\documentclass[12pt]{iopart}
\usepackage{iopams}
\usepackage{graphicx}

\begin{document}

\title{Optical vortices of slow light using tripod scheme}

\author{J~Ruseckas, A~Mekys and G~Juzeli{\=u}nas}

\address{Institute of Theoretical Physics and Astronomy, Vilnius University,
A.~Go\v{s}tauto 12, Vilnius 01108, Lithuania}

\eads{\mailto{julius.ruseckas@tfai.vu.lt}, \mailto{algirdas.mekys@ff.vu.lt}, \mailto{gediminas.juzeliunas@tfai.vu.lt}}

\begin{abstract}
We consider propagation, storing and retrieval of slow light (probe beam) in a
resonant atomic medium illuminated by two control laser beams of larger
intensity. The probe and two control beams act on atoms in a tripod
configuration of the light-matter coupling. The first control beam is allowed to
have an orbital angular momentum (OAM). Application of the second vortex-free
control laser ensures the adiabatic (lossles) propagation of the probe beam at
the vortex core where the intensity of the first control laser goes to zero.
Storing and release of the probe beam is
accomplished by switching off and on the control laser beams leading to the
transfer of the optical vortex from the first control beam to the regenerated
probe field. A part of the stored probe beam remains frozen in the medium in the 
form of atomic spin excitations, the number of which increases with increasing
the intensity of the second control laser. We analyse such losses in the
regenerated probe beam and provide conditions for the optical vortex of the
control beam to be transferred efficiently to the restored probe beam.
\end{abstract}

\vspace{2pc}

\submitto{\JO}

\maketitle

\section{Introduction}

Over the last decade there has been a great deal of activities in slow
\cite{Hau-99,Kash99,Budker99,Novikova07PRL,Davidson09NP}, stored
\cite{Fleischhauer-00,Liu-01,Phill2001,Juzeliunas-02,Scully02PRL,Yanik05PRA,Gor05EPJD,Lukin05Nature,Ite07PRA,Hau07Nature,Bloch09PRL,Hau09PRL,Beil10PRA}
and stationary
\cite{Stationary-light-03--09,Stationary-light2,Stationary-light4,Stationary-ligth3,Stationary-ligth5,Otterbach10PRL,Unanyan10}
light. It was demonstated that a resonant
weak pulse of light (to be referred to as the probe light) can propagate as
slow as several of tens of meters per second \cite{Hau-99} in an atomic medium
driven by a stronger (control) laser beam. The application of the control laser
beam makes the resonant and opaque medium transparent for the probe beam due
the electromagnetically induced transparence (EIT)
\cite{Arimondo-96,Harris-97,Scully-97,Lukin-03,Fleischhauer-05}, both beams
making a $\Lambda$ configuration of the atom-light coupling.
The EIT can be used not only to slow down dramatically the light pulses
\cite{Hau-99,Kash99,Budker99,Novikova07PRL,Davidson09NP},
but also to store them
\cite{Liu-01,Phill2001,Lukin05Nature,Hau07Nature,Bloch09PRL,Hau09PRL,Beil10PRA}
in atomic gases.
The storage and release of the probe pulses has been accomplished
\cite{Liu-01,Phill2001,Lukin05Nature,Hau07Nature,Bloch09PRL,Hau09PRL,Beil10PRA}
by switching off and on the control laser \cite{Fleischhauer-00}. The slow and
storred light can be applied to reversible quantum memories
\cite{Fleischhauer-00,Juzeliunas-02,Scully02PRL,Lukin05Nature,Lukin-03,Fleischhauer-05,Appel08PRL,Honda08PRL,Akiba09NJP} and, in the case of moving media
\cite{Leonhardt00PRL,Ohberg02PRAR,Fleischhauer-Gong-PRL02,Juz-moving,Artoni03PRA,Zimmer-Fleischhauer-PRL04,Padgett-06,Ruseckas-07},
also to rotational sensing devices.

The orbital angular momentum (OAM) \cite{Allen-99,Allen-03} provides additional
possibilities in manipulating the slow light. The optical OAM represents a new
degree of freedom which can be exploited in the quantum computation and quantum
information storage \cite{Allen-03}. Most of the previous studies on the vortex
slow light have concentrated on situations in which the incident probe beam
carries an OAM
\cite{Ruost04PRL,Davidson07PRL,Ruseckas-07,Yelin08PRA,Moretti09PRA}. 

Here we consider another situation in which the incident probe beam does not have an
optical vortex, yet it gains the OAM when retrieved. For this purpose one of
the control laser beam is assumed to have an optical vortex during either the
storage or retrieval of the slow light. The intensity of such a control beam
goes to zero at the vortex core, potentially leading to the absorption losses
of the probe beam in this area. To avoid the losses an extra control laser
without an optical vortex is used. This makes a more complex tripod scheme of
the atom-light coupling
\cite{Unanian,Knight-02,Rebic04PRA,Petrosyan04PRA,Yelin04PRA,Rus05,Mazets05PRA,Zaremba06OC,Zaremba07PRA,Zaremba07OC,Garva07OC,RusecMekJuz08}.
The total intensity of the control lasers is then non-zero at the vortex core of the
first control laser preventing the absorption losses. Subsequently we analyze additional losses
appearing because a part of the stored probe beam remains frozen in the atomic cloud
in the form of spin excitations during the exchange of the optical vortex between the control laser and
the regenerated slow light. A number of such frozen excitations increases with increasing
the intensity of the second control laser. We provide conditions for the optical vortex of the
control beam to be transferred efficiently to the restored probe beam.

\section{Equations for the probe beam}

In this Section we shall derive general equations for the propagation of the
probe beam in an ensemble of tripod atoms. In doing so we shall not make use of
the usual slowly-varying amplitude approximation
\cite{Zaremba06OC,Zaremba07OC,Zaremba07PRA}
for the spatial coordinates in the direction perpendicular to the light
propagation. This will make it possible to analyse situations where the control
and/or probe beams have the OAM.

\subsection{Coupled equations for the probe beam and atoms}

\begin{figure}
{\centering
\includegraphics[width=0.6\textwidth]{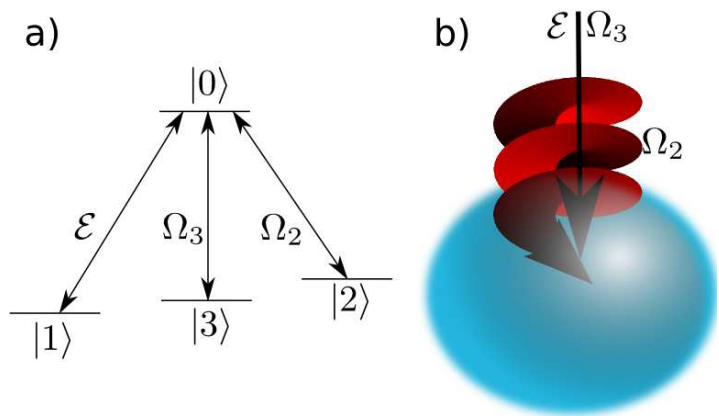}\par
}
\caption{Light induced transitions in the cloud of cold atoms involving a
tripod configuration of the internal atomic states.}
\label{fig:fig1}
\end{figure}

Let us consider an ensemble of tripod-type atoms characterised by three
hyperfine ground levels $1$, $2$, and $3$, as well as an electronic excited
level $0$ (figure~\ref{fig:fig1}). The translational motion of atoms is
represented by a four component column operator $\Psi(\bi{r})$, where the
components $\Psi_1(\bi{r},t)$, $\Psi_2(\bi{r},t)$,
$\Psi_3(\bi{r},t)$, and $\Psi_0(\bi{r},t)$ are the field operators
describing to the center of mass motion in the four internal atomic states. The
quantum nature of the atoms comprising the medium will determine whether these
field operators obey Bose-Einstein or Fermi-Dirac commutation relations. The
atoms interact with three light fields in a tripod configuration of the
atom-light coupling \cite{Unanian,Rus05,Zaremba06OC,Zaremba07PRA,Zaremba07OC}.
Specifically, two strong classical control lasers induce transitions
$|2\rangle\rightarrow|0\rangle$, $|3\rangle\rightarrow|0\rangle$ and a weaker
quantum probe field drives a transition $|1\rangle\rightarrow|0\rangle$, as
shown in figure~\ref{fig:fig1}.

The electric field of the probe beam is 
\begin{equation}
\bi{E}(\bi{r},t)=\hat{\bi{e}}\sqrt{\frac{\hbar\omega}{
2\varepsilon_0}}\mathcal{E}(\bi{r},t)\rme^{-\rmi\omega t}+\mathrm{H.c.}\,,
\end{equation}
where $\omega=ck$ is the central frequency of the probe photons,
$\bi{k}=\hat{\bi{z}}k$ is the wave vector, and
$\hat{\bi{e}}\bot\hat{\bi{z}}$ is the unit polarization vector. This
field can be considered to be either a quantum operator or a classical
variable. We have chosen the dimensions of the electric field amplitude
$\mathcal{E}$ to be such that its squared modulus represents the number density
of probe photons. The probe field is assumed to be quasi-monochromatic, so the
amplitude $\mathcal{E}(\bi{r},t)$ changes little over the time of an
optical cycle and thus obeys the following equation: 
\begin{equation}
\left(\partial_t-\rmi\frac{c^2}{2\omega}\nabla^2-\rmi\frac{\omega}{2}\right)
\mathcal{E}=\rmi g\Phi_1^{\dag}\Phi_0\,,
\label{eq:electric}
\end{equation}
where have introduce the slowly varying atomic field operators
$\Phi_1=\Psi_1\rme^{\rmi\omega_1t}$,
$\Phi_2=\Psi_2\rme^{\rmi(\omega_1+\omega-\omega_{c2})t}$,
$\Phi_3=\Psi_3\rme^{\rmi(\omega_1+\omega-\omega_{c3})t}$,
$\Phi_0=\Psi_0\rme^{\rmi(\omega_1+\omega)t}$, with $\hbar\omega_1$ being the energy
of the atomic ground state $1$, $\omega_{c2}$ and $\omega_{c3}$ being the
frequencies of the control fields. The parameter
$g=\mu\sqrt{\omega/2\varepsilon_0\hbar}$ featured in equation~\eref{eq:electric}
characterizes the strength of coupling of the probe field with the atoms ,
$\mu$ being the dipole moment of the atomic transition
$|1\rangle\rightarrow|0\rangle$. Note that, unlike in the usual treatment of
slow light, we have retained the second-order derivative $\nabla^2$ in the
equation of motion~\eref{eq:electric}. This allows one to account for the fast
changes of $\mathcal{E}$ in a direction perpendicular to the wave vector
$\bi{k}$, i.e., in the $xy$ plane. Therefore our analysis can be applied to
the twisted beams of light $\mathcal{E}(\bi{r},t)\sim\exp(\rmi l\varphi)$
carrying an OAM $\hbar l$ per photon.

In the following we shall make use of a semi-classical picture in which both
the electromagnetic and matter field operators are replaced by $c$ numbers. The
equations for the matter fields are then 
\begin{eqnarray}
\rmi\partial_t\Phi_1 = -g\mathcal{E}^*\Phi_0 \label{eq:phi1} \\
\rmi\partial_t\Phi_0 =
(\omega_{01}-\rmi\gamma)\Phi_0-\Omega_{\mathrm{c}2}\Phi_2
-\Omega_{\mathrm{c}3}\Phi_3-g\mathcal{E}\Phi_1\label{eq:phi0}\\
\rmi\partial_t\Phi_2 = \omega_{21}\Phi_2
-\Omega_{\mathrm{c}2}^*\Phi_0 \label{eq:phi2} \\
\rmi\partial_t\Phi_3 = \omega_{31}\Phi_3
-\Omega_{\mathrm{c}3}^*\Phi_0\,,
\label{eq:phi3}
\end{eqnarray}
Here $\Omega_{\mathrm{c}2}$ and $\Omega_{\mathrm{c}3}$ are the Rabi frequencies of control lasers
driving the transitions $|2\rangle\rightarrow|0\rangle$ and
$|3\rangle\rightarrow|0\rangle$; $\gamma$ is the decay rate of the excited
level; $\omega_{21}=\omega_2-\omega_1+\omega_{c2}-\omega$ and
$\omega_{31}=\omega_3-\omega_1+\omega_{c3}-\omega$ are frequencies of the
electronic detuning from the two-photon resonances,
$\omega_{01}=\omega_0-\omega_1-\omega$ is the frequency of the electronic
detuning from the one-photon resonance. In writing
equations~\eref{eq:phi1}--\eref{eq:phi3} we used the rotating wave approximation
for the atom-light coupling, and neglected the terms containing atomic mass
$m$, since from now on we are not interested in the effects due to the atomic
motion: The terms containing atomic mass $m$ would be important in
equations~\eref{eq:phi1}--\eref{eq:phi3} for the description of the light-dragging
effects
\cite{Fleischhauer-Gong-PRL02,Juz-moving,Zimmer-Fleischhauer-PRL04,Padgett-06,Ruseckas-07}.
Furthermore we assumed that the trapping potentials for an atom in
the electronic state $j$ ($j=1,2,3,0$) are position independent and hence can
be omitted.


\subsection{Equation for the of probe beam}

To analyze the atomic dynamics it is convenient to introduce the bright and
dark states of the atomic center of mass motion 
\begin{eqnarray}
\Phi_{\mathrm{B}} =\Omega_{\mathrm{c}}^{-1}(\Omega_{\mathrm{c}2}\Phi_{2}
+\Omega_{\mathrm{c}3}\Phi_{3})\,,\label{eq:B-state}\\
\Phi_{\mathrm{D}} =\Omega_{\mathrm{c}}^{-1}(\Omega_{\mathrm{c}3}^{*}\Phi_{2}
-\Omega_{\mathrm{c}2}^{*}\Phi_{3})\,,
\end{eqnarray}
where 
\begin{equation}
\Omega_{\mathrm{c}}=\sqrt{|\Omega_{\mathrm{c}2}|^2+|\Omega_{\mathrm{c}3}|^2}
\end{equation}
is the total Rabi frequency. It is the wave-function $\Phi_{\mathrm{B}}$ which is featured
in the equation of motion~\eref{eq:phi0} for the excited state wave-function.
On the other hand, the dark state is not coupled directly to the excited level
$0$. The original fields $\Phi_2$ and $\Phi_3$ can be expressed through
$\Phi_{\mathrm{B}}$ and $\Phi_{\mathrm{D}}$ as 
\begin{eqnarray}
\Phi_2 = \Omega_{\mathrm{c}}^{-1}(\Omega_{\mathrm{c}2}^*\Phi_B
+\Omega_{\mathrm{c}3}\Phi_D)\,, \label{eq:phi2-DB} \\
\Phi_3 = \Omega_{\mathrm{c}}^{-1}(\Omega_{\mathrm{c}3}^*\Phi_B
-\Omega_{\mathrm{c}2}\Phi_D)\,.\label{eq:phi3-DB}
\end{eqnarray}
Initially the atoms are in the ground level $1$ and the Rabi frequency of the
probe field is considered to be much smaller than $\Omega_{\mathrm{c}}$. Consequently one
can neglect the last term in equation~\eref{eq:phi1} that causes depletion of the
ground level $1$, giving $i\partial_t\Phi_1=0$. If the atoms in the internal
ground-state $1$ form a BEC, its wave function $\Phi_1=\sqrt{n}\exp(iS_1)$
represents an incident variable determined by the atomic density $n$ and the
condensate phase $S_1$. The latter phase will not play an important role in our
subsequent analysis, since we are not interested in the effects due to the
condensate dynamics.

Suppose the control and probe beams are tuned close to the two-photon
resonance. Application of such laser beams cause electromagnetically induced
transparency (EIT) in which the transitions $|2\rangle\rightarrow|0\rangle$,
$|3\rangle\rightarrow|0\rangle$, and $|1\rangle\rightarrow|0\rangle$ interfere
destructively preventing population of the excited state $0$. The adiabatic
approximation is obtained neglecting the excited state population in
equation~\eref{eq:phi0}, giving 
\begin{equation}
\Phi_{\mathrm{B}}=-g\frac{\Phi_1}{\Omega_{\mathrm{c}}}\mathcal{E}\,.
\label{eq:phiB-electric}
\end{equation}
When the adiabatic approximation is valid, the bright and dark states are
coupled weakly. Therefore, combining equations~\eref{eq:phi2}), \eref{eq:phi3} and
neglecting the dark state contribution $\Phi_D$, we arrive at the equation for
the bright state wave-function 
\begin{equation}
\rmi\partial_t\Phi_{\mathrm{B}}=\delta\Phi_{\mathrm{B}}-\Omega_{\mathrm{c}}\Phi_0\,,
\label{eq:bright}
\end{equation}
where
\begin{equation}
\delta=\omega_{21}|\xi_2|^2+\omega_{31}|\xi_3|^2-\rmi(\xi_2\partial_t\xi_2^*
+\xi_3\partial_t\xi_3^*)
\label{eq:U}
\end{equation}
is the two-photon frequency mismatch, with 
\begin{equation}
\xi_2=\Omega_{\mathrm{c}2}/\Omega_{\mathrm{c}}\,,\qquad\xi_3=
\Omega_{\mathrm{c}3}/\Omega_{\mathrm{c}}\,.
\label{eq:zeta-2--3}
\end{equation}
The derivation of equation~\eref{eq:bright} can be found in more details in our
earlier work \cite{Juz-effective-05} on the light-induced gauge potentials for
the $\Lambda$ type atoms. Equation~\eref{eq:bright} relates $\Phi_0$ to the
bright state $\Phi_{\mathrm{B}}$ as 
\begin{equation}
\Phi_0=\Omega_{\mathrm{c}}^{-1}(-\rmi\partial_t+\delta)\Phi_{\mathrm{B}}\,.
\label{eq:phi0-phiB}
\end{equation}
Finally, equations~\eref{eq:electric}, \eref{eq:phiB-electric} and
\eref{eq:phi0-phiB} provide a closed equation for the electric field amplitude
$\mathcal{E}$: 
\begin{equation}
\left(\partial_t-\rmi\frac{c^2}{2\omega}\nabla^2-\rmi\frac{\omega}{2}\right)
\mathcal{E}=-\rmi\frac{g^2\Phi_1^*}{\Omega_{\mathrm{c}}}(-\rmi\partial_t+\delta)
\frac{\Phi_1}{\Omega_{\mathrm{c}}}\mathcal{E}\,.
\label{eq:electric-2}
\end{equation}
This equation applies a wide variety of phenomena. In particular it can be used
to model light storage by introducing time dependence in $\Omega_{\mathrm{c}}$ or light
dragging due to spatial variation of $\Omega_{\mathrm{c}}$ or $\Phi_1$.

\subsection{Non-adiabatic corrections}

\label{sub:regeneration}In the next Section we shall consider the releasing of
the stored light. For this we should include non-adiabatic corrections to the
equation of motion~\eref{eq:electric-2}. This can be done in the following
way. From equation~\eref{eq:phi0} expressing the bright state $\Phi_{\mathrm{B}}$ and
substituting equation~\eref{eq:phi0-phiB} for $\Phi_0$ we get 
\begin{equation}
\Phi_{\mathrm{B}}=-\frac{g\mathcal{E}}{\Omega_{\mathrm{c}}}\Phi_1
+\Omega_{\mathrm{c}}^{-1}(-\rmi\partial_t+\omega_{01}-\rmi\gamma)
\Omega_{\mathrm{c}}^{-1}(-\rmi\partial_t+\delta)\Phi_{\mathrm{B}}\,.
\end{equation}
The term with the decay rate $\gamma$ is larger than other non-adiabatic
corrections in the above equation. Keeping in the non-adiabatic corrections
only the terms proportional to the decay rate $\gamma$ and neglecting, for
simplicity, the two-photon detuning $\delta$ we obtain 
\begin{equation}
\Phi_{\mathrm{B}}\approx-\frac{g\mathcal{E}}{\Omega_{\mathrm{c}}}\Phi_1
-\frac{\gamma}{\Omega_{\mathrm{c}}^2}\partial_t\Phi_{\mathrm{B}}\,.
\label{eq:phiB-aprox}
\end{equation}
The solution of this equation, assuming that the control beam is switched on
suddenly at $t=0$ an then changes slowly during the characteristic relaxation
time $\gamma/\Omega_{\mathrm{c}}^2$, is 
\begin{equation}
\Phi_{\mathrm{B}}=\Phi_{\mathrm{B}}(0)\rme^{-\frac{\Omega_c^2}{\gamma}t}
-\frac{g\mathcal{E}}{\Omega_{\mathrm{c}}}\Phi_1
\left(1-\rme^{-\frac{\Omega_{\mathrm{c}}^2}{\gamma}t}\right)\,.
\label{eq:psiB-result}
\end{equation}
Using equations~\eref{eq:phi0-phiB} and \eref{eq:psiB-result}, the equation for
the electric field \eref{eq:electric} takes the form
\begin{eqnarray}
\fl\left(\partial_t-\rmi\frac{c^2}{2\omega}\nabla^2-\rmi\frac{\omega}{2}\right)
\mathcal{E} = 
-\rmi\frac{g^2\Phi_1^*}{\Omega_{\mathrm{c}}}(-\rmi\partial_t+\delta)
\frac{\Phi_1}{\Omega_{\mathrm{c}}}\mathcal{E}
\left(1-\rme^{-\frac{\Omega_{\mathrm{c}}^2}{\gamma}t}\right)\nonumber \\
+\rmi\frac{g\Phi_1^*}{\Omega_{\mathrm{c}}}(-\rmi\partial_t+\delta)\Phi_{\mathrm{B}}(0)
\rme^{-\frac{\Omega_{\mathrm{c}}^2}{\gamma}t}\,.
\label{eq:e-generation-memeber}
\end{eqnarray}
At $t=0$ the probe field is off, and the information on the previously stored
probe beam being contained in the atomic coherence $\Phi_{\mathrm{B}}(0)$. The
regeneration of the probe beam is described by the second term on the
r.h.s.\ of equation~\eref{eq:e-generation-memeber} representing the source for the electric
field. Retaining only the temporal derivatives in
equation~\eref{eq:e-generation-memeber}, we get the equation describing the
generation of the electric field:
\begin{equation}
\left[1+\frac{g^2n}{\Omega_{\mathrm{c}}^2}
\left(1-\rme^{-\frac{\Omega_{\mathrm{c}}^2}{\gamma}t}\right)\right]
\partial_t\mathcal{E}=
-\frac{g\Phi_1^*}{\gamma}[g\Phi_1\mathcal{E}+\Omega_{\mathrm{c}}\Phi_{\mathrm{B}}(0)]
\rme^{-\frac{\Omega_{\mathrm{c}}^2}{\gamma}t}\,,
\end{equation}
with the initial condition $\mathcal{E}(0)=0$ at $t=0$. For time in access of
the relaxation time $\gamma/\Omega_{\mathrm{c}}^2$ the regenerated probe field evolves to
a steady-state value complying with the adiabatic condition
\eref{eq:phiB-electric}
\begin{equation}
\mathcal{E}=-\frac{\Omega_{\mathrm{c}}}{g\Phi_1}\Phi_{\mathrm{B}}(0)\,.
\label{eq:E-regenerated}
\end{equation}
In this way, the regenerated electric field $\mathcal{E}$ is indeed determined
by the initial atomic coherence $\Phi_{\mathrm{B}}(0)$. The subsequent evolution of the
probe field is described by the adiabatic equation of motion
\eref{eq:electric-2} containing both the temporal and spatial derivatives
subject to the initial condition \eref{eq:E-regenerated}.

\subsection{Co-propagating control and probe beams}

Suppose the probe and control beams co-propagate:
$\mathcal{E}(\bi{r},t)=\tilde{\mathcal{E}}(\bi{r},t)\rme^{\rmi kz}$,
$\Omega_{\mathrm{c}2}(\bi{r},t)=\Omega_2(\bi{r},t)\rme^{\rmi k_{\mathrm{c}2}z}$,
$\Omega_{\mathrm{c}3}(\bi{r},t)=\Omega_3(\bi{r},t)\rme^{\rmi k_{\mathrm{c}3}z}$,
where $k_{\mathrm{c}2}$ and $k_{\mathrm{c}3}$ are the wave numbers of the control beams.
For paraxial beams the
amplitudes $\tilde{\mathcal{E}}(\bi{r},t)$, $\Omega_2(\bi{r},t)$ and
$\Omega_3(\bi{r},t)$ depend weakly in the propagation direction $z$
compared to the variation of the exponential factors. Equation
\eref{eq:electric-2} for the probe field takes then the form 
\begin{equation}
\partial_t\tilde{\mathcal{E}}+v_{\mathrm{g}}
\left[\frac{\partial}{\partial z}+\left(\frac{1}{v_{\mathrm{g}}}
-\frac{1}{c}\right)\rmi\delta-\rmi\frac{1}{2k}\nabla_{\bot}^2\right]
\tilde{\mathcal{E}}=\left(1-\frac{v_{\mathrm{g}}}{c}\right)
\frac{\partial_t\Omega_{{\mathrm{c}}}}{\Omega_{{\mathrm{c}}}}\tilde{\mathcal{E}}\,,
\label{eq:electric-3}
\end{equation}
where we have replaced $\nabla^2$ by its transverse part
$\nabla_{\bot}^2=\partial^2/\partial x^2+\partial^2/\partial y^2$ because of
the paraxial approximation. Here 
\begin{equation}
v_{\mathrm{g}}=c\left(1+\frac{g^2n}{\Omega_{\mathrm{c}}^2}\right)^{-1}
\end{equation}
is the radiative group velocity. The term with spatial derivative
$\partial/\partial z$ in equation~\eref{eq:electric-3} describes the radiative
propagation along the $z$ axis with the group velocity $v_{\mathrm{g}}$.

\section{Storing and releasing the light}

The probe beam $\mathcal{E}^{(\mathrm{s})}$ enters an atomic medium which is illuminated
by two control beams characterized by the Rabi frequencies
$\Omega_{\mathrm{c}2}^{(\mathrm{s})}$ and $\Omega_{\mathrm{c}3}^{(\mathrm{s})}$,
where the index
$(\mathrm{s})$ refer to the stage of storing the light. At the boundary the
probe beam is converted into a polariton propagating slowly in the medium with
the velocity $v_{\mathrm{g}}^{(\mathrm{s})}\ll c$. At certain time $t=t^{(\mathrm{s})}$
the whole probe
pulse enters the atomic medium and is contained in it. Since the atomic
population is created exclusively by the incident light field, the atomic
dark-state $\Phi_{\mathrm{D}}$ is not populated and, according to
equation~\eref{eq:phiB-electric}, the bright-state is
\begin{equation}
\Phi_{\mathrm{B}}^{(\mathrm{s})}(t^{(\mathrm{s})})=
-g\frac{\Phi_1}{\Omega_{\mathrm{c}}^{(\mathrm{s})}}
\mathcal{E}^{(\mathrm{s})}(t^{(\mathrm{s})})\,.
\end{equation}
Equations~\eref{eq:phi2-DB} and \eref{eq:phi3-DB} give the atomic fields:
\begin{equation}
\Phi_2^{(\mathrm{s})}=\xi_{\mathrm{c}2}^{(\mathrm{s})*}
\Phi_{\mathrm{B}}^{(\mathrm{s})}\,,\qquad\Phi_3^{(\mathrm{s})}=
\xi_{\mathrm{c}3}^{(\mathrm{s})*}\Phi_{\mathrm{B}}^{(\mathrm{s})}\,.
\label{eq:Psi2-3-s}
\end{equation}
To store the slow light, both control fields are switched off at $t=t^{(\mathrm{s})}$ in
such a way that the ratios $\xi_{\mathrm{c}2}^{(\mathrm{s})}$ and
$\xi_{\mathrm{c}3}^{(\mathrm{s})}$ remain constant whereas
$\Omega_{\mathrm{c}}^{(\mathrm{s})}\rightarrow 0$.
The stored atomic coherences no longer have the radiative group velocity and
thus are trapped in the medium. To restore the slow light propagation, the
control fields are switched on again at $t=t^{(\mathrm{r})}$ with relative Rabi
frequencies $\xi_{\mathrm{c}2}^{(\mathrm{r})}$ and
$\xi_{\mathrm{c}3}^{(\mathrm{r})}$. The latter can differ from
the original ones $\xi_{\mathrm{c}2}^{(\mathrm{s})}$ and
$\xi_{\mathrm{c}3}^{(\mathrm{s})}$, so
the dark-state $\Phi_{\mathrm{D}}$ can now be populated. Shortly after the beginnig of the
release of light (at $t=t^{(\mathrm{r})}$) the generated electric field reaches a steady
state value, as described in the Subsection~\ref{sub:regeneration}.
Equation~\eref{eq:E-regenerated} yields the restored probe field:
\begin{equation}
\mathcal{E}^{(\mathrm{r})}(t^{(\mathrm{r})})=
-\frac{\Omega_{\mathrm{c}}^{(\mathrm{r})}}{g\Phi_1}(\xi_{\mathrm{c}2}^{(\mathrm{r})}
\xi_{\mathrm{c}2}^{(\mathrm{s})*}
+\xi_{\mathrm{c}3}^{(\mathrm{r})}\xi_{\mathrm{c}3}^{(\mathrm{s})*})
\Phi_{\mathrm{B}}^{(\mathrm{s})}(t^{(\mathrm{s})})\,,
\label{eq:electric-general-restore}
\end{equation}
where equations~\eref{eq:B-state}, \eref{eq:zeta-2--3} and \eref{eq:Psi2-3-s}
were used to relate the bright state $\Phi_{\mathrm{B}}^{(\mathrm{r})}(t^{(\mathrm{r})})$
of the restoring stage to the stored one $\Phi_{\mathrm{B}}^{(\mathrm{s})}(t^{(\mathrm{s})})$ .

Since a typical length of the atomic cloud is not much larger than the length
of the laser pulse, we will assume that the stored and restored electric fields
do not change significantly during their propagation inside of the atomic
cloud. Furthermore we will assume that the control beams are abruptly switched
off during the storing stage and then switched on in the same way when they are
restored. Using equations~\eref{eq:phiB-electric} and
\eref{eq:electric-general-restore}, the restored field may be written as:
\begin{equation}
\mathcal{E}^{(\mathrm{r})}=
\frac{\Omega_{\mathrm{c}}^{(\mathrm{r})}}{\Omega_{\mathrm{c}}^{(\mathrm{s})}}
(\xi_{\mathrm{c}2}^{(\mathrm{r})}\xi_{\mathrm{c}2}^{(\mathrm{s})*}
+\xi_{\mathrm{c}3}^{(\mathrm{r})}\xi_{\mathrm{c}3}^{(\mathrm{s})*})
\mathcal{E}^{(\mathrm{s})}\,.
\label{eq:electric-restored-2}
\end{equation}

\subsection{Control beams with the same spatial behaviour}

Supposed first that the Rabi frequencies of the restored control beams are
proportional to the original ones:
$\Omega_{\mathrm{c}2}^{(\mathrm{r})}=a\Omega_{\mathrm{c}2}^{(\mathrm{s})}$ and
$\Omega_{\mathrm{c}3}^{(\mathrm{r})}=a\Omega_{\mathrm{c}3}^{(\mathrm{s})}$.
For slow light this implies
that $\xi_{\mathrm{c}2}^{(\mathrm{r})}=\xi_{\mathrm{c}2}^{(\mathrm{s})}$,
$\xi_{\mathrm{c}3}^{(r)}=\xi_{\mathrm{c}3}^{(\mathrm{s})}$. Since
$\xi_{\mathrm{c}2}\xi_{\mathrm{c}2}^*+\xi_{\mathrm{c}3}\xi_{\mathrm{c}3}^*=1$,
equation~\eref{eq:electric-restored-2}
yields the following result for the regenerated electric field:
\begin{equation}
\mathcal{E}^{(\mathrm{r})}(t^{(\mathrm{r})})=
-\frac{\Omega_{\mathrm{c}}^{(\mathrm{r})}}{g\Phi_1}
\Phi_{\mathrm{B}}^{(\mathrm{s})}(t^{(\mathrm{s})})\,.
\end{equation}
The above relationship represents the initial condition for the subsequent
propagation of the probe beam $\mathcal{E}$ governed by the equation of
motion \eref{eq:electric-3}. The regenerated electric field is seen to acquire
the phase from the bright polariton at its storage stage and the amplitude of
$\mathcal{E}$ is modulated according to $\Omega_{\mathrm{c}}^{(\mathrm{r})}$
at the release stage.

\subsection{Transfer of optical vortex at the retrieval of the probe beam}

Suppose now that initially we have a $\Lambda$ system with only a single
control field: $\xi_{\mathrm{c}3}^{(\mathrm{s})}=0$ and hence
$|\xi_{\mathrm{c}2}^{(\mathrm{s})}|=1$. On the other hand, a tripod system is used in
the retrieval stage where generally both $\xi_{\mathrm{c}2}^{(\mathrm{r})}$ and
$\xi_{\mathrm{c}3}^{(\mathrm{r})}$
are non-zero. In that case equation~\eref{eq:electric-general-restore} provides the
following result for the regenerated electric field:
\begin{equation}
\mathcal{E}^{(\mathrm{r})}(t^{(\mathrm{r})})=
-\frac{\Omega_{\mathrm{c}2}^{(\mathrm{r})}}{g\Phi_1}\xi_{\mathrm{c}2}^{(s)*}
\Phi_{\mathrm{B}}^{(\mathrm{s})}(t^{(\mathrm{s})})\,.
\label{eq:electric-lamda-case}
\end{equation}
The equation \eref{eq:electric-lamda-case} represents the initial condition
for the subsequent propagation of the probe beam $\mathcal{E}$ governed by the
equation of motion \eref{eq:electric-3} in the medium.

If the second control beam has an optical vortex at the restoring stage
$\Omega_{\mathrm{c}2}^{(\mathrm{r})}\sim \rme^{\rmi\ell\varphi}$,
the regenerated electric field
$\mathcal{E}^{(\mathrm{r})}\sim \rme^{\rmi\ell\varphi}$ acquires the same phase, as one can
see from equation~\eref{eq:electric-lamda-case}. This means the restoring control
beam transfers its optical vortex to the regenerated electric field
$\mathcal{E}^{(\mathrm{r})}$. In the $\Lambda$ scheme it is not allowed to have an
optical vortex for the control beam due to adiabaticity violation at the center
of the vortex. Using a tripod scheme for the regeneration lifts up this
restriction. The probe beam may itself carry a vortex at the beginning of the
storage \cite{Ruost04PRL}. Subsequently the vortex is stored onto the
atomic bright state $\Phi_{\mathrm{B}}^{(\mathrm{s})}$ and then transferred back to the probe beam
after the control beam $\Omega_{\mathrm{c}2}^{(\mathrm{r})}$ is turned on. In
that case the phase of the restored vortex in the probe beam is defined by the product
$\Omega_{\mathrm{c}2}^{(\mathrm{r})}\Phi_{\mathrm{B}}^{(\mathrm{s})}$ in which
both $\Omega_{\mathrm{c}2}^{(\mathrm{r})}$ and
$\Phi_{\mathrm{B}}^{(\mathrm{s})}$ may carry vortices. If these vortices have opposite winding
numbers, they cancel each other leading to zero vorticity in the regenerated
probe beam.

Let us take the restoring control laser $\Omega_{\mathrm{c}2}^{(\mathrm{r})}$ to be the first
order Laguerre-Gaussian (LG) beam:
$\Omega_{\mathrm{c}2}^{(\mathrm{r})}=A\tilde{\rho}\rme^{\rmi\varphi}\exp(-\tilde{\rho}^2/
\sigma_{\mathrm{r}}^2)$,
where $\tilde{\rho}=\rho/\lambda$ is a dimensionless cylindrical radius,
$\lambda=2\pi/k$ being the optical wave-length. On the other hand, the control
beam is assumed to be the zero-order LG beam during the storage stage involving
a $\Lambda$ system:
$\Omega_{\mathrm{c}2}^{(\mathrm{s})}=a^{-1}A\exp(-\tilde{\rho}^2/\sigma_{\mathrm{s}}^2)$,
where $a$ determines a relative amplitude of the control fields
$\Omega_{\mathrm{c}2}^{(\mathrm{r})}$ and $\Omega_{\mathrm{c}2}^{(\mathrm{s})}$,
$\sigma_{\mathrm{r}}$ and $\sigma_{\mathrm{s}}$ being
their dimensionless widths. This provides the following regenerated probe field
\eref{eq:electric-lamda-case}
\begin{equation}
\mathcal{E}^{(\mathrm{r})}=a\tilde{\rho}\rme^{\rmi\varphi}\exp[-\tilde{\rho}^2
(\sigma_{\mathrm{r}}^{-2}-\sigma_{\mathrm{s}}^{-2})]\mathcal{E}^{(\mathrm{s})}\,.
\label{eq:E-r-lambda-storing}
\end{equation}

\subsection{Transfer of the optical vortex during the storage of slow light}

Suppose now that initially we have a tripod system for the storage where
generally both $\xi_{\mathrm{c}2}^{(\mathrm{s})}$ and
$\xi_{\mathrm{c}3}^{(\mathrm{s})}$ are non-zero. On the other
hand, a $\Lambda$ system is used in the retrieval stage with only one control
field, i.e.\ $\xi_{\mathrm{c}3}^{(\mathrm{r})}=0$ and hence
$|\xi_{\mathrm{c}2}^{(\mathrm{r})}|=1$. In that case
equation~\eref{eq:electric-general-restore} leads to the same
equation~\eref{eq:electric-lamda-case} for the regenerated electric field. Yet now
it is the storing control beam that has an optical vortex
$\Omega_{\mathrm{c}2}^{(\mathrm{s})}\sim \rme^{\rmi\ell\varphi}$. Subsequently
the vortex is transfered to the regenerated probe field in the phase conjugated
form: $\mathcal{E}^{(\mathrm{r})}\sim \rme^{-\rmi\ell\phi}$.

Suppose that the control lasers are the first and zero order LG beams at the
storage stage: 
\begin{equation}
\Omega_{\mathrm{c}2}^{(\mathrm{s})}=A\tilde{\rho}\rme^{\rmi\varphi}
\exp(-\tilde{\rho}^2/\sigma_{\mathrm{s}}^2)\,,\quad\Omega_{\mathrm{c}3}^{(\mathrm{s})}
=bA\exp(-\tilde{\rho}^2/\sigma_{\mathrm{s}}^2)\,,
\end{equation}
where the parameter $b$ determines the relative amplitude of the additional
control laser. On the other hand, the control beam is assumed to be the
zero-order LG beam at the retrieval stage involing the $\Lambda$ scheme:
$\Omega_{\mathrm{c}2}^{(\mathrm{r})}=aA\exp(-\tilde{\rho}^2/\sigma_{\mathrm{r}}^2)$.
Thus one arrives at the following regenerated probe field containing the phase
conjugated vortex
\begin{equation}
\mathcal{E}^{(\mathrm{r})}=\frac{a}{\tilde{\rho}^2+b^2}\tilde{\rho}\rme^{-\rmi\varphi}
\exp[-\tilde{\rho}^2(\sigma_{\mathrm{r}}^{-2}-\sigma_{\mathrm{s}}^{-2})]
\mathcal{E}^{(\mathrm{s})}\,.
\label{eq:E-r-lambda-release}
\end{equation}

\subsection{Energy losses}

\begin{figure}
{\centering
\includegraphics[width=0.6\textwidth]{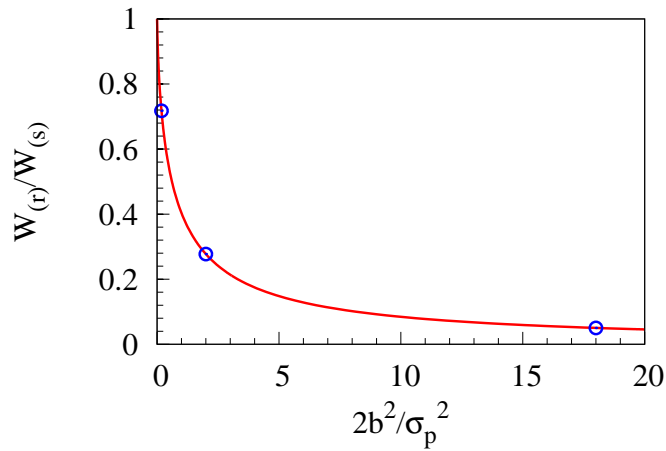}\par
}
\caption{The dependence of loses on the relative amplitude ratio between the
control beams $b$ calculated using equation~\eref{eq:losesFinal}. Three
circles indicate three cases corresponding to $\sigma_p=10$ and $b=3$, $10$ and $30$.}
\label{fig:loses}
\end{figure}

The proposed shemes for transferring a vortex from the control laser beam to the
regenerated probe beam avoid non-adiabaticit (absporption) losses at the center
of the vortex due to application of an additional control laser. Yet there are
another kind of losses in the energy of the regenerated probe beam, because a
part of the stored probe beam remains frozen in the medium in the form of atomic
spin excitations. Let us estimate those losses. Suppose that the incident pulse
of the probe electric field $\mathcal{E}^{(\mathrm{s})}$ has a Gaussian
intensity profile with a width $\sigma_{\mathrm{p}}$ along its transversal
coordinate $\tilde{\rho}$ and a length $L_{\mathrm{vac}}$ in the propagation
direction $z$. The total electric field energy entering the cloud of cold atoms
is
\begin{equation}
W_{(\mathrm{s})}=2\pi\int_0^{+\infty}\tilde{\rho}\rmd\tilde{\rho}
\int_0^{L_{\mathrm{vac}}}\rmd z\,|\mathcal{E}^{(\mathrm{s})}|^2=
\frac{\pi}{2}L_{\mathrm{vac}}|\mathcal{E}_0|^2\sigma_{\mathrm{p}}^2\,.
\end{equation}
The electric probe field at the retrieval stage is related to the stored on in a
different way for two cases of the storage and retrieval considered above.

In the case of lambda storage and tripod retrieval, regenerated electric field
is described by the equation~\eref{eq:E-r-lambda-storing}. The length of the
regenerated pulse is $L_{(\mathrm{r})}$. This length is related to the length
$L_{\mathrm{vac}}$ of the initial pulse via the equation
$L_{(\mathrm{r})}=L_{\mathrm{vac}}v_{\mathrm{g}}^{(\mathrm{s})}/v_{\mathrm{g}}^{(\mathrm{r})}$.
Since the group velocity $v_{\mathrm{g}}^{(\mathrm{r})}$ in the retrieval stage
depends transversal coordinate $\tilde{\rho}$, the length $L_{(\mathrm{r})}$ is
also $\tilde{\rho}$ dependent. The total energy of electric field leaving the
cloud is
\begin{eqnarray}
W_{(\mathrm{r})} 
& = 2\pi\int_{0}^{+\infty}\tilde{\rho}\rmd\tilde{\rho}\int_{0}^{L_{(\mathrm{r})}}
\rmd z\,|\mathcal{E}^{(\mathrm{r})}|^{2} \nonumber \\
& = \pi L_{\mathrm{vac}}|\mathcal{E}_{0}|^{2}\int_{0}^{+\infty}
\frac{x\exp[-2x(\sigma_{\mathrm{r}}^{-2}+\sigma_{\mathrm{p}}^{-2})]}{x
\exp(-2x/\sigma_{\mathrm{r}}^{2})+b^{2}\exp(-2x/\sigma_{\mathrm{r}3}^{2})}\rmd x\,.
\end{eqnarray}
Here $\sigma_{\mathrm{r}3}$ is the width of the second control beam at the
retrieval stage
$\Omega_{\mathrm{c}3}^{(\mathrm{r})}=bA\exp(-\tilde{\rho}^2/\sigma_{\mathrm{r}3}^2)$.
The energy losses may be found comparing the energies of the pulses before and
after the interaction with the cloud. For simplicity suppose that
$\sigma_{\mathrm{r}}=\sigma_{\mathrm{r}3}$. Then
\begin{eqnarray}
\frac{W_{(\mathrm{r})}}{W_{(\mathrm{s})}} 
& = \int_{0}^{+\infty}\frac{\rme^{-y}}{y+2b^{2}/\sigma_{\mathrm{p}}^{2}}y\rmd y\nonumber \\
& = 1+\frac{2b^{2}}{\sigma_{\mathrm{p}}^{2}}
\exp\left(\frac{2b^{2}}{\sigma_{\mathrm{p}}^{2}}\right)
\mathop{\mathrm{Ei}}\left(-\frac{2b^{2}}{\sigma_{\mathrm{p}}^{2}}\right)\,.
\label{eq:losesFinal}
\end{eqnarray}
If the intensity of the second control beam is large, $b\gg\sigma_{\mathrm{p}}$,
the ratio of the energies is
$W_{(\mathrm{r})}/W_{(\mathrm{s})}\approx\sigma_{\mathrm{p}}^2/(2b^2)$.

In the case of tripod storiage and lambda retrieval the restored electric field
profile is described by equation~\eref{eq:E-r-lambda-release}. Nevetheless the
loses are found the same as in the previous case \eref{eq:losesFinal}. Several
values of $W_{(\mathrm{r})}/W_{(\mathrm{s})}$ are given in
figure~\ref{fig:loses}. When the amplitude ratio of the control beams $b$
increases, the loses are seen to increase.

\section{Concluding remarks}

We have considered propagation, storing and retrieval of slow light (probe beam)
in a resonant atomic medium illuminated by two control laser beams of larger
intensity. The probe and two control beams act on atoms in a tripod
configuration of the light-matter coupling in which three hyperfine atomic
ground states and one excited state are involved. The first control beam is
allowed to have an orbital angular momentum (OAM). Application of the second
vortex-free control laser ensures the adiabatic (lossles) propagation of the
probe beam at the vortex core where the intensity of the first control laser
goes to zero. Using the adiabatic approximation we have derived the equation of
motion for the probe beam and analysed it in the case where one of the control
beams has an optical vortex.  

Storing and release of the probe beam is accomplished by switching off and on
the control laser beams leading to the transfer of the optical vortex from the
first control beam to the regenerated probe field. A part of the stored probe
beam remains frozen in the atomic cloud in a form of spin excitations, a
number of which increases with increasing the intensity of the second control
laser. We have analysed such losses in the regenerated probe beam and provided
conditions for the optical vortex of the control beam to be transferred
efficiently to the restored probe beam.

We have investigated in detail two cases of storing and retrieval of optical
vortices onto the atomic medium. In the first case the $\Lambda$ scheme is used
for the storage, whereas the tripod setup is exploited for the retrieval. In
such a situation the vortex can be transferred efficiently and without a
distortion from the restoring control beam to the regenerated probe beam. In the
second case the tripod system is used for the storing and the $\Lambda$ system
is employed for the retrieval. The vortex is then transferred from the storing
control beam to the regenerated probe in a phase conjugated form, so the
regenerated probe beam acquires an opposite vorticity. The regenerated beam is
then distorted and becomes narrower as compared to the storing beam. Thus it
experiences a larger diffraction spreading in the subsequent propagation in the
medium.

\ack

The authors acknowledge the support by the Research Council of Lithuania
(Grants No. MOS-2/2010 and VP1-3.1-\v{S}MM-01-V-01-001)

\end{document}